# ASSESSMENT TWINS: A PROTOCOL FOR AI-VULNERABLE SUMMATIVE ASSESSMENT

Jasper Roe [1*], Mike Perkins [2], Louie Giray [3],

[1] Durham University, United Kingdom

[2] British University Vietnam, Vietnam

[3] Mapúa University, Philippines

* Corresponding Author: jasper.j.roe@durham.ac.uk

## Abstract

Generative Artificial Intelligence (GenAI) is reshaping higher education and raising pressing concerns about the integrity and validity of higher education assessment. While assessment redesign is increasingly seen as a necessity, there is a relative lack of literature detailing what such redesign may entail. In this paper, we introduce *assessment twins* as an accessible approach for redesigning assessment tasks to enhance validity. We use Messick's unified validity framework to systematically map the ways in which GenAI threaten content, structural, consequential, generalisability, and external validity. Following this, we define assessment twins as two deliberately linked components that address the same learning outcomes through different modes of evidence, scheduled closely together to allow for cross-verification and assurance of learning.

We argue that the twin approach helps mitigate validity threats by triangulating evidence across complementary formats, such as pairing essays with oral defences, group discussions, or practical demonstrations. We highlight several advantages: preservation of established assessment formats, reduction of reliance on surveillance technologies, and flexible use across cohort sizes. To guide implementation, we propose a three-step design process: identifying vulnerabilities, aligning outcomes, selecting complementary tasks, and developing interdependent marking schemes. We also acknowledge the challenges, including resource intensity, equity concerns, and the need for empirical validation. Nonetheless, we contend that assessment twins represent a validity-focused response to GenAI that prioritises pedagogy while supporting meaningful student learning outcomes.

**Keywords:** assessment twin, generative artificial intelligence, assessment design, assessment validity, higher education

# Introduction

Generative Artificial Intelligence (GenAI) remains a crucial area of research in assessment practice in higher education. Since the public release of advanced GenAI models, concerns regarding academic integrity have risen sharply, as the production of stylistically and grammatically coherent text is now widely available (Giray, 2024a; Perkins, 2023). Given that many forms of higher educational assessment have traditionally relied on formats such as the take-home essay, written research project, or pre-prepared presentation, which can now be completed by GenAI, the potential effects on assurance of learning are troubling.

The paradigm-rupturing nature of GenAI has also prompted deeper reflection on the purpose of assessment and the role of higher education itself (Bannister et al., 2025; Giray et al., 2024). Current discourse demonstrates significant uncertainty as to whether GenAI will ultimately help or hinder assessment practices, with significant concern remaining around student use of GenAI in completing tasks. Critics argue that such reliance may lead to reduced learner agency (Roe & Perkins, 2024), dependency, and erosion of critical thinking skills (Giray, 2025; Gonsalves, 2024). These contradictory perspectives underscore the unsettled nature of current educational debates: GenAI presents both opportunities and risks, inviting innovation and threatening core pedagogical aims.

In this paper, we outline a practice for redesigning assessments in higher education in light of the capabilities of developing GenAI models. Specifically, we propose the concept of *assessment twins* for GenAI-vulnerable tasks. In practice, this entails pairing each task that may be susceptible to GenAI assistance or completion (such as a take-home essay) with a second, less vulnerable task that assesses the same outcomes. This design has several advantages. First, it strengthens assessment validity by generating confirmatory data on the same set of learning outcomes. Second, it preserves the pedagogical value of established assessment formats that should not be discarded entirely and enables judicious, authentic, and appropriate engagement with GenAI, which has been named as a core principle of assessment redesign in the AI era (TEQSA, 2025). Third, it aligns with broader calls for educators to emphasise collaborative, in-person, and multimodal forms of assessment in the age of GenAI (Rudolph et al., 2023) and the use of multiple, inclusive, and contextualised methods of assessment to form "trustworthy judgements about student learning" (TEQSA, 2025, p1).

The remainder of this paper is organised as follows: We begin by reviewing the current literature on GenAI and assessment validity. We then introduce the concept of an assessment twin in depth and explain how it enhances multiple strands of validity. Finally, we offer guidelines for practical implementation and conclude with a discussion of the limitations of the twin framework.

# Literature

The use of technology-assisted platforms to aid in written academic work (with associated impacts on assessment validity) predates GenAI (Prentice & Kinden, 2018; Roe & Perkins, 2022). However, the advanced capabilities of GenAI to produce extended works has led to a focus on GenAI-assisted plagiarism, or 'Aigiarism' (Khalaf, 2025). As a result, it is challenging to identify whether a students' work is their own. This compromises validity as it becomes impossible to identify whether students have met the required standards for a course (Dawson et al., 2024), and thus fails to provide assurance of learning. Despite the introduction of GenAI tools into the educational landscape several years ago, no clear answer has emerged to resolve this so-called 'wicked problem' (Corbin, Bearman, et al., 2025). AI detection was rapidly promoted as a potential remedy, but studies have shown that these technologies do not work well enough to make informed decisions on student usage (Chaka, 2023; Perkins, Roe, et al., 2024; Weber-Wulff et al., 2023), and so detecting GenAI use in assessments is now "all but impossible" (TEQSA, 2025, p2). Furthermore, such surveillance-focused responses may impact the relational dimension of assessment. Carless (2009) highlighted this point, suggesting that trust must be developed between students and institutions for effective assessment reform.

Outside of detection, other strategies have also been posited. These include relying less on surveillance technologies and more on providing student support (Luo, 2024), embedding AI literacy into higher education curricula (Foung et al., 2024), incorporating self-reflection tasks (Combrinck & Loubser, 2025), abandoning certain assessment types (Kofinas et al., 2025), and embedding contextual learning elements into assessment (Gonsalves, 2025). Essien et al. (2024) contend that offering clear ethical guidelines may prevent GenAI misuse, while Cotton et al. (2024) suggest that a mixture of approaches, including educating students, requiring multiple draft submissions, using detection tools, and closely monitoring student work, are all potentially effective strategies.

Frameworks and systems for fostering assessment and audit and redesign have also been developed, including the Assessment-GenAI Susceptibility Rubric PANDORA (Bannister et al., 2025), AI Assessment Scale (Perkins et al., 2025; Perkins, Furze, et al., 2024), 'traffic light' systems to communicate acceptable GenAI usage (University of Leeds, 2025), and 'lanes' for secured and unsecured assessment with and without access to GenAI (Bridgeman et al., 2024). It has however, been argued that approaches which only communicate guidelines without accompanying structural changes (discursive changes) are inadequate for dealing with GenAI in assessment (Corbin et al., 2025).

While assessment redesign is not a silver bullet for addressing the issue of GenAI in assessment in higher education, we contend that it is a valuable method for enhancing assessment validity. At the same time, we contend that existing forms of assessment (such as take-home essays, portfolios, and unsupervised, authentic pieces of work) still have legitimate value and a place in summative assessment protocols. This is at the core of the philosophy behind creating assessment twins.

We frame our understanding of validity through Messick's (1989, 1993) work on validity. Traditional conceptions of assessment validity are classified into three types: content, criterion-related, and construct validity. Messick (1989, 1993) challenged this assertion, proposing a unified model in which construct validity is the overriding framework under which all other validity aspects are subsumed. According to this model, validity can be defined as an evaluative judgement on the extent to which evidence supports the appropriateness, meaningfulness, and usefulness of assessment results. Messick did not explicitly label his framework as consisting of six strands, but later works (Shaw & Crisp, 2012) have drawn on Messick to frame six sources of validity evidence: content, substantive, structural, generalisability, external, and consequential. Each of these contributes to the overall construct validity, as shown in Table 1.

**Table 1: Shaw and Crisp's (2012) Six-Strands of Validity, based on Messick (1989, 1993)**

| Validity Strand | Definition |
| --- | --- |
| Content | Relates to the representativeness and relevance of the content. |
| Substantive | Relates to the justifications and theoretical basis for the consistency of the assessment, how comparable the underlying cognitive processes are vis-à-vis the assessment and performance in practice. |
| Structural | Relates to how reliable the procedures for assigning scores and scoring processes are. |
| Consequential | Regards the consequences of the assessment for the person who is taking the assessment. |
| Generalisability | Asks to what degree can score properties or interpretations be widened and generalised in different contexts. |
| External | Relates to the relationship between assessment scores and scores on other assessments which measure the same thing. |

## Assessment Twins

The premise behind an assessment twin is that when one form of assessment is more vulnerable to GenAI completion, pairing it with a complementary task that is less vulnerable provides greater assessment validity and a clearer representation of assessment performance. Consequently, we define assessment twins as two deliberately designed, interdependent assessment components that (a) address the *same* intended learning outcomes, (b) require *different modes of evidence or production*, and (c) are scheduled so that performance on each component can be cross-checked to mitigate a known vulnerability (e.g. GenAI completion, impersonation), thereby enhancing validity compared to either component considered alone.

A twin strategy does not require educators to abandon established assessment types such as essays or reports, which can be pedagogically valuable. In contrast, an assessment twin acknowledges the role of these assessments but seeks to enhance their validity by gathering additional evidence. The twin approach builds on existing, long-established assessment practices, such as the oral viva voce, which is commonly associated with thesis defences in postgraduate assessment. We also foresee an assessment twin protocol as suited to summative assessment, in which the objective is to judge learning (Bennett, 2011; Crisp, 2012) or certify achievement (Craddock & Mathias, 2009).

Creating an assessment twin is distinct from the simple process of pairing a written essay with a traditional oral viva voce. Notably, a twin task for a GenAI-susceptible assessment could take multiple formats, including a group interview or peer discussion, a timed in-class test, or the production of a physical artefact. The underlying principle is one of complementary modes of assessment which promote authenticity, creating a system of checks and balances where inconsistencies in understanding, proficiency, or competency come to the surface.

A further benefit of twinned assessments is their flexibility. Twin elements can be both low-tech (in-class discussions, oral defences) or high-tech (for example, creating an in-class concept map as a group using AI tools). The twin approach can also be applied to small classes of a few individuals or larger groups.

## How Do Assessment Twins Enhance Validity?

Messick's (1989) six-strand approach provides a lens for analysing how GenAI disrupts traditional assumptions regarding assessment validity (see Table 2). Each of these strands of validity is now exposed to new, uneven pressures in GenAI enabled educational contexts. Tasks which previously aligned with certain constructs may now be exposed to shortcuts, or score interpretation may no longer be reliable.

The impact of GenAI on these dimensions of validity is not singular; rather, GenAI may affect several different strands of validity simultaneously. In Table 2, we map the ways in which these six strands of validity are disrupted by the attributes of GenAI models and propose ways in which validity could be strengthened through the redevelopment of assessment through a twin process.

**Table 2: Mapping GenAI Validity Threats and Assessment Twinning Responses to Strands of Validity**

| Validity Type | GenAI Threat | Assessment Twin Strategy | Validity is Enhanced by |
|---|---|---|---|
| Content Validity | Learners submit work to an assessment designed to evaluate knowledge on an issue using GenAI models, effectively bypassing the learning themselves. | Twin take-home assessments with in-person discussion of key concepts. | Confirmation of understanding as it relates to learning outcomes. |
| Substantive Validity | Learners bypass specific cognitive processes or synthesis techniques through GenAI usage (e.g. Using OpenAI's deep research to conduct a literature review). | Have learners explain or document their cognitive processes. | Evidence of engagement with required cognitive process. For example, a demonstration of a core skill. |
| Structural Validity | GenAI produced content receives high assessment scores, undermining score validity. | Link scores across assessment elements, e.g., cap scores on a written task if oral explanation is poor | Cross-verifying range of performance to maintain score reliability. |
| Consequential Validity | The assessment encourages a surface-level approach and induces dependency on GenAI tool usage. This detracts from learner agency and autonomy in the long term, thus is a negative consequence induced by GenAI vulnerability | Incorporate a twin which has a metacognitive element (i.e. an oral self-assessment) and ask students to incorporate a reflection on their use of GenAI. | Promotes self regulation, critical awareness of GenAI impacts, and stimulated learner reflection |
| Generalisability | Learners perform well only in environments where they have technology access, but fail to replicate performance | Improve generalisability of results by assessing the same set of learning outcomes in low and | Assessing the ability to apply knowledge in GenAI and non- |

|  | when they do not have GenAI access | high technology contexts, with restrictions on AI use where necessary. | GenAI enabled contexts |
|---|---|---|---|
| **External Validity** | Assessment scores are not reliable indicators of real-world performance | Include practice-based or scenario tasks for human-confirmed elements | Triangulation of applied and non-applied skills. |

In summary, each of these elements of overall construct validity may be enhanced by adopting a twin approach in the assessment strategy.

# Practical Design for Twin Assessments

While we provide conceptual evidence for the twin concept to enhance assessment validity, it is important that this approach is grounded in practical implementation strategies. As a novel framework, there are no existing empirical cases of a twin strategy in action. However, in proposing assessment twins, we foresee that the protocol would be best implemented through an iterative audit and development process. We propose a three- step process here, beginning with the identification of a vulnerability, followed by the consideration of learning outcomes and development of a complementary assessment, followed by the creation of a marking framework, prior to pilot testing the assessment. The proposed steps are outlined below.

**Step 1: Identifying Assessment Vulnerability**

The first step towards creating assessment twins is to identify whether the existing assessment(s) are threatened by the capabilities of GenAI. This requires the assessor to have a threshold level of AI literacy, for example, by understanding the strengths and limitations of current GenAI models and what they can and cannot do. Broadly speaking, if the assessment outcomes are threatened by the production of high-quality GenAI output with little to no human input, then there is a strong argument that the validity of the assessment is challenged. Tools such as the PANDORA rubric (Bannister et al., 2025) may be of value in this part of the process. Additional considerations are the context in which the assessment takes place: if students are able to undertake the assessment remotely, without human observation, supervision, or invigilation, then there is a greater likelihood that the validity of the assessment will be threatened. Additionally, the ease with which GenAI content can be differentiated from human work may be a deciding factor. An art project undertaken using canvas and oil paints, for example, would meet the criteria of being achievable or completable remotely, without supervision, yet it would not be meaningfully vulnerable to GenAI completion. If assessments fulfil most, or all of these criteria, then requiring a twin assessment may help to maintain validity.

**Step 2: Consider Learning Outcomes and Choose a Twin Assessment**

The following step is to define and explicate the learning outcomes that the assessment is required to measure. This includes any competencies, skills, or knowledge which are required to pass the assessment. Assessment twins should not be two disconnected tasks, and the value of the approach lies in the fact that both components should map onto the same set of learning outcomes in a complementary way. By focusing on the intended learning outcomes, educators can identify which dimensions of learning are most likely to be compromised by GenAI. There are multiple ways that a twin could be designed. The exact format of an assessment twin depends on the learning context, institutional requirements and resource constraints, and the nature of the subject being assessed. Complementary modes to traditional written assessments could include a real-time demonstration, for example an oral explanation, group discussion, or question and answer session. However, in resource-limited contexts with large student cohorts, this may not be a feasible option. In these cases, forms of peer-assessment could be explored, for example team-based assessment. The twin should be less vulnerable to GenAI completion, while retaining measurement of the intended learning outcomes.

By way of example, if a learning outcome relates to being able to critically evaluate source material, then a written essay may be useful to demonstrate clearly structured arguments, while an oral discussion or video recording of a reflection on the work may help verify the students' reasoning process. A second example could be the application of knowledge in practice: in this case, a simulation of an authentic task or in-class problem based task could be combined with a secondary written report.

**Step 3: Develop a Marking Framework**

A key principle behind an assessment twin is the interdependence between the two tasks. This does not mean that a grade or weighting is assigned for one component and the other (i.e. a 50% weighting on a pre-prepared presentation and a 50% weighting on an interview). The marking approach to the assessment needs careful consideration. This could include a confirmatory aspect, for example the performance in the second assessment is required to confirm the performance in the first assessment (i.e. it is a 'yes/no'). A threshold may also be established, to suggest a minimum performance on the twin to validate the GenAI vulnerable assessment.

We recognise that there are contexts in summative higher education assessments where assessment twins may not be appropriate, as shown in Table 3.

**Table 3: When (not) to use twins**

| Use a twin when… | Do not twin when… |
| --- | --- |
| <ul><li>A task is pedagogically rich but AI-susceptible (essay, take-home coding, design brief).</li><li>The same LOs can be evidenced via a second, lower-risk mode (supervised discussion, rapid in-class derivation, oral walk-through, process log).</li><li>Institutional constraints rule out full invigilation but allow other authenticity checks.</li></ul> | <ul><li>Outcomes differ across the two tasks: A single redesign (e.g., authentic, supervised studio task) would already be valid and manageable.</li><li>Workload or equity concerns (e.g., time-intensive viva voces for large cohorts) cannot be mitigated with scalable alternatives.</li><li>High-stakes, single-sitting exams are feasible and already secure</li></ul> |

# Strategies for Implementing a Twin Assessment Design

Twin assessment design offers an approach to maintaining validity in a world in which GenAI tools are widely available. However, we recognise that implementing assessment restructuring such as a twin strategy is easier said than done, and heavily context specific. In this section, we explore some of the implementation challenges associated with a twin assessment practice.

One of the fundamental issues that we anticipate in terms of using twin assessments in higher education is the resource-intensiveness of providing additional, in-person confirmatory tasks (such as oral viva voces) which may not be possible in the context of large class sizes.

We also recognise that there are contexts in which an assessment twin approach will not be appropriate. Therefore, we suggest that assessment twinning be chosen only in specific circumstances as discussed in Table 3. For example, if an existing assessment is pedagogically valuable yet GenAI vulnerable, this is the most important criterion for implementing a twin strategy. Even assessments potentially vulnerable to GenAI may still retain pedagogical value, and instructors may still wish to retain these as part of a formal assessment task, rather than changing them to a formative assessment or learning activity.

Further to this, assessment twin strategies are suited to institutions which can support the resources required for implementation, and in contexts where student learning is enhanced from multimodal assessment. In contrast, when dealing with resource-limited contexts or extremely large class sizes, then we would argue that a twin approach may still be a possible, but non-optimal solution. In these cases, a complete redesign of the overall assessment strategy using an established framework (such as the AIAS) is more likely to yield results in enhancing validity.

**Twin assessment strategies for different cohort sizes**

In line with the above, we recognise that the administration of assessments and assessment redesign must be focused on the realities of cohort sizes. We categorise these as small groups (between 5 and 25 students), medium groups (25 – 75 students) and large groups (75 and above).

*Small Group Assessment Twin Strategies (5 – 25 students)*

In a smaller group size, there is greater potential for the instructor to interact personally with each student and develop a relationship, over time and through formative assessment potentially understanding the learners' position, capabilities, and areas for development. This lends itself to relatively resource intensive assessment twin design.

Furthermore, a smaller group size requires fewer logistical considerations. In this context, a GenAI vulnerable assessment could be paired with an individual oral examination or viva voce, a small group discussion with rotating student facilitators, a peer review session, or an individual consultation. In terms of implementation, twin components may be scheduled during learning hours or classes, or in dedicated assessment time. In this context, a twin assessment approach will provide quality validity evidence.

*Medium Group Assessment Twin Strategies (25 – 75 students)*

As the size of a cohort increases, so too do the resource requirements for designing and delivering twin assessments. A strategy to mitigate this is to incorporate group assessment formats. Examples of an assessment that could be twinned with a GenAI-vulnerable assessment element include peer-group presentation, larger simultaneous group discussions in which the instructor briefly spends time with each group, or a poster presentation event. If possible, incorporating multiple assessors may make this approach more viable.

*Large Group Assessment Twin Strategies (75+ students)*

A large group which requires multiple forms of assessment poses resource constraints and significant challenges to implementing a twinned assessment practice, but this is still viable with some caveats. Bearing in mind that validity is always a claim, rather than an absolute (Dawson, 2020), there are still benefits to a twinned approach in terms of providing validity evidence. Examples of a twinned assessment strategy that would be effective in enhancing validity for such a group could include a random sampling approach, in which a percentage of students are selected for a detailed twin assessment, or peer group discussion sessions with multiple assessors (if practical). Video submissions may be vulnerable to technological manipulation such as deepfakes (Runyon, 2025) yet still could provide another significant data point in collaboration with other, more significantly GenAI-vulnerable tasks (such as a take-home, written assessment). In-person examinations of course remain an important and secure form of assessment, and could be part of a twinning approach, if the same outcomes as the GenAI vulnerable assessment are being assessed.

# Limitations and Future Research

**Implementation and Scalability**

The most pressing challenge of implementing assessment twins in higher education is resource intensity: this approach requires faculty time, administrative coordination, and institutional support. For large student cohorts, scalability becomes especially problematic. While we suggest solutions, such as random sampling, these risk undermining the very validity the twin strategy is designed to protect.

We also face complex administrative barriers, including scheduling logistics, maintaining consistent scoring across multiple assessors, and reconciling grades when twin components produce conflicting outcomes. Even if we can overcome these hurdles, quality assurance may remain uncertain. We must therefore acknowledge that while assessment twins have value, they may not yet be practical in some educational contexts, especially those already facing resource constraints.

**Equity and Accessibility**

We must also confront serious questions of fairness. As assessment research warns, design choices that overlook inclusivity can unintentionally deepen inequities (Lynam & Cachia, 2018). Assessment twins may disadvantage students with diverse communication styles, social anxiety, or cultural backgrounds that make oral examinations and group discussions especially challenging. Students with disabilities may face additional barriers if accommodations are not considered across both components.

Meanwhile, increased assessment load risks placing unequal pressure on students who juggle family responsibilities, employment, or limited study time. We also recognize that language barriers may unfairly affect international students or those for whom English is not a first language, particularly in oral formats. Perhaps most concerning, if we frame the approach primarily around catching GenAI misuse, we risk fostering a surveillance mentality that positions students as potential cheaters rather than learners (Giray, 2024b; Dawson, 2020), undermining the trust essential to education. Without deliberate attention to inclusive design, we may unintentionally create more inequitable learning environments rather than fairer ones.

**Empirical Limitations**

We must acknowledge that to date, no empirical data exists on the effectiveness of this approach. While the framework assumes that inconsistencies between twin components provide assurance of learning by identifying discrepancies between performance in each assessment task, we recognize that such inconsistencies could just as easily reflect legitimate factors such as anxiety, uneven skill development, or differences in comfort with assessment formats (Struyven et al., 2005), and different assessment formats may capture different learning outcomes rather than equivalent ones (Shaw & Crisp, 2012). Elshall and Badir (2025) have called for hybrid approaches that combine traditional methods with AI-assisted projects, but we note that such models remain largely unexplored. Still, despite these significant limitations, we believe assessment twins have utility. Engaging in a twin process forces us to grapple with urgent questions about validity, fairness, and trust in an era of rapidly evolving GenAI tools. In this sense, assessment twins should not be seen as a perfect or final solution, but as a bold and necessary experiment.

**Future Research Directions**

Future research needs to prioritize the empirical validation of the assessment twins framework through systematic investigation across multiple educational contexts, building on established approaches in assessment research (Boud et al., 2018). This is not simply about testing whether the framework works. Our key priorities should include controlled pilot studies in small-scale settings to test effectiveness in providing triangulation through multiple data points, and mixed-methods research that captures both quantitative outcomes (such as grade correlations) and qualitative experiences (including student stress levels and faculty workload); and longitudinal studies that track whether twin assessments actually enhance learning and academic performance. We should also engage in comparative research that evaluates assessment twins alongside alternative strategies such as authentic assessment reforms (Crisp, 2012) or AI-integrated pedagogies (Foung et al., 2024). Just as importantly, we must explore whether twin performance links meaningfully to real-world competencies, giving us a stronger basis for claims of predictive validity. By pursuing these lines of inquiry, we not only fill the current evidence gap but also build a more credible and resilient framework.

# Conclusion

In this paper, we have proposed *assessment twins* as a response to the challenges created by Generative AI in higher education. Our framework aims to help ensure validity of assessment, while preserving the pedagogical value of established tasks. By pairing two assessments that address the same learning outcomes through different modes of evidence, we provide opportunities for cross-verification and generate stronger claims about assurance of learning.

We argue that validity is strengthened when multiple tasks converge on the same outcome, and we emphasise that the adaptability of twins across different cohort sizes makes this approach widely relevant. At the same time, we recognise the limitations. Assessment twins demand extra resources, thoughtful workload management, and inclusive design to avoid inequities. Without careful planning, risks such as stress, inefficiency, or superficial adoption may undermine the potential of a twin approach. We therefore call for further research, experimentation, and refinement to test and improve the model. Despite these challenges, we contend that this twin assessment strategy may be one of many methods that can support in enhancing the validity of assessments and supporting learning in the new GenAI era.

Assessment in higher education must move beyond surveillance-driven practices toward a model where multiple assessment modes become complementary learning opportunities (Giray et al., 2025). In this sense, the assessment twins concept aligns with calls for a post-assessment future attentive to power and inequality, where assessment redesigns respond to systemic shifts rather than merely policing tools (Perkins & Roe, 2025).

**Acknowledgements**

We are grateful for the ideas contributed by Leon Furze and Thomas Corbin in the initial development phases of this piece.